# MAC design for WiFi infrastructure networks: a game-theoretic approach

Ilenia Tinnirello, Laura Giarré and Giovanni Neglia

***Abstract*—In WiFi networks, mobile nodes compete for accessing a shared channel by means of a random access protocol called Distributed Coordination Function (DCF). Although this protocol is in principle fair, since all the stations have the same probability to transmit on the channel, it has been shown that unfair behaviors may emerge in actual networking scenarios because of non-standard configurations of the nodes. Due to the proliferation of open source drivers and programmable cards, enabling an easy customization of the channel access policies, we propose a game-theoretic analysis of random access schemes. Assuming that each node is rational and implements a best response strategy, we show that efficient equilibria conditions can be reached when stations are interested in both uploading and downloading traffic. More interesting, these equilibria are reached when all the stations play the same strategy, thus guaranteeing a fair resource sharing. When stations are interested in upload traffic only, we also propose a mechanism design, based on an artificial dropping of layer-2 acknowledgments, to force desired equilibria. Finally, we propose and evaluate some simple DCF extensions for practically implementing our theoretical findings.**

## I. Introduction

The problem of resource sharing in WiFi networks [1], [2], is addressed by the Distributed Coordination Function (DCF), which is a Medium Access Control (MAC) protocol based on the paradigm of carrier sense multiple access with collision avoidance (CSMA/CA). The basic idea of the protocol is very simple: sensing the channel before transmitting, and waiting for a random backoff time when the channel is sensed busy. This random delay, introduced for preventing collisions among waiting stations, is slotted for efficiency reason and extracted in a range called contention window. Standard DCF assumes that the contention window is set to a minimum value ($CW_{min}$) at the first transmission attempt and is doubled up to a maximum value ($CW_{max}$) after each transmission failure.

The distributed DCF protocol is in principle fair, because the contention window settings $CW_{min}$ and $CW_{max}$ are homogeneous among the stations, thus ensuring that each node receives in long term the same number of access opportunities. Nevertheless, some unexpected behaviors have been recognized as a consequence of non-standard settings of the contention windows. In fact, stations employing lower contention windows gain probabilistically a higher number of transmission opportunities, at the expense of compliant stations. These settings can be performed by the card manufacturers, as recognized in [3], or by the end users thanks to the availability of open-source drivers.

Another problem specific to infrastructure networks is given by the repartition between uplink and downlink resources. Infrastructure networks are characterized by a star topology, which connects multiple mobile nodes to a common station called Access Point (AP). On one side, mobile stations can upload traffic to the AP, which is connected to external networks (e.g. to the internet); on the other side, they can download traffic from the external networks through the AP. Since the AP contends as a normal station to the channel, its channel access probability is the same of other mobile stations. This implies that the AP aggregated throughput, i.e. the downlink bandwidth, is equal to the throughput perceived by all the other stations, thus resulting in a per-station downlink bandwidth much lower than the uplink one [4]. Indeed, recent extensions of DCF [5] (namely, the EDCA protocol) allow the AP to set heterogeneous contention windows among the stations to give priority to downlink throughput or to delay-sensitive traffic. Thus, nowadays nodes can adapt their contention windows according to the values signaled by the AP for each traffic class.

I. Tinnirello is with DIEET, Università di Palermo, 90128 Palermo, Italy ilenia.tinnirello@tti.unipa.it

L. Giarré is with DIAS, Università di Palermo, 90128 Palermo, Italy giarre@unipa.it

G. Neglia is with Équipe Projet INRIA Maestro, 2004 route des Lucioles, 06902 Sophia Antipolis, France, giovanni.neglia@inria.fr

However, there is the risk to exploit this adaptation in a selfish manner, for example by using a contention window value of an higher priority class [6].

These considerations motivate a game theoretical analysis of DCF, in order to propose some protocol extensions able to cope with the current resource sharing problems. The problem can be formulated as a non cooperative game, in which the contending stations act as the players of the game. When stations work in saturation conditions, i.e. they always have a packet available in the transmission buffer, DCF can be modeled as a slotted access protocol, while station behavior can be summarized in terms of per-slot access probability [7].

Let $\tau_i$ be the per-slot access probability representing the access strategy of a generic station $i$. The channel access game can be formulated by considering: $n$ players, the set of strategies $\tau = (\tau_i, i = 1, \ldots, n)$ in $[0, 1]^n$, and the station payoff $(J_1, J_2, \ldots, J_n)$, that can be defined according to the network and application scenario [8]. Previous studies have mainly considered that such a utility is given by the node upload throughput performance [9]. In [10], it has been shown that a utility function equal to the upload throughput may lead to an inefficient Nash equilibrium in which stations transmit in every channel slot (i.e. play $\tau = 1$). This situation creates a resource collapse, because all stations transmit simultaneously thus destroying all packet transmissions. More complex utility functions combining upload throughput and costs related to collision rates [10], [11], [12] or to energy consumptions [13], [14] lead to different equilibria, but they implicitly assume that all the nodes have the same energy constraints or collision costs. Other utility functions, which lead to efficient equilibria, appear less natural [15], because they are not related to physical parameters, such as throughput.

In this paper, we show that efficient equilibria conditions can be naturally reached when stations are interested in both upload and download traffic. Since the utility of each station depends not only on its throughput but also on the AP throughput, no station is motivated to transmit continuously. We propose a game theoretic analysis of DCF in infrastructure networks, when all the stations have a desired ratio between uplink and downlink throughput. Assuming that each station tunes its access probability according to a best response strategy, extending our preliminary results in [16] and in [17], we derive Nash equilibria and Pareto optimal conditions as a function of the network scenario. We also define a mechanism design scheme, in which the AP plays the role of arbitrator to improve the global performance of the network, by forcing desired equlibria conditions. We propose to extend current DCF operation by implementing our theoretical best response strategies. To this purpose, we develop some channel monitoring functionalities (similarly to [18], [19]), devised to estimate the network status and to run-time drive the strategy adaptations. Finally, we evaluate the effectiveness of our scheme by means of extensive simulations.

The rest of the paper is organized as follows. In section II we carry the game theoretic analysis and we find the Nash equilibria and the Pareto Optimal solutions; in section III we design the DCF extension; in section IV we show the MAC scheme implementation and the performance evaluation trough simulations; finally we drew some conclusive remarks in section V.

## II. CONTENTION-BASED CHANNEL ACCESS: A GAME THEORETIC ANALYSIS

We assume that all the stations try permanently to transmit on the channel, i.e. they work in saturation conditions. In fact, we have verified that non-saturated stations affect the performance of saturated stations only marginally and regardless of their contention windows. When all stations are saturated, it has been shown [20] that DCF can be accurately approximated as a persistent slotted access protocol, because packet transmissions can be originated only at given time instants. Figure 1 shows an example of DCF as a slotted protocol. After a busy time, stations A and B defer their transmissions by extracting a random slotted delay (respectively, 4 and 8 slots). Since station A expires its backoff first, it acquires the right to transmit on the channel. The next transmission, which results in a collision, is performed again after an integer number of backoff slots from the end of the previous channel activity. Therefore, the channel time can be divided into slots of uneven duration delimited by potential transmission instants. In a generic channel slot, each station has approximately a fixed probability $\tau$ to transmit (in the example, $1/5$ for station A and $1/10$ for station B)), which depends on the average backoff values.

### A. Station strategies

Let $n$ be the number of saturated contending stations. We assume that each station $i$ is rational, and can arbitrarily choose its channel access probability $\tau_i$ in $[0, 1]$. This choice can be readily implemented by tuning opportunistically the minimum and the maximum values of the contention windows. By observing that $\tau_i = 1/(1 + E[W]/2)$, where $E[W]$ is the average contention window used by station, a solution is to set $CW^i_{min} = CW^i_{max} = 2/\tau_i - 2$.

The set of all the strategies in the network is then $[0, 1]^n$. We define an *outcome* of the game a specific set of strategies taken by the players, then a vector $\boldsymbol{\tau} = (\tau_1, \tau_2, \cdots, \tau_n) \in [0, 1]^n$. We define that an outcome is *homogeneous* whenever all the stations play the same strategy, i.e. $\boldsymbol{\tau} = (\tau, \tau, ...\tau)$.

Being all the player requirements homogeneous, the equilibria we are going to define are invariant to player permutations. For this reason we define two outcomes $\boldsymbol{\tau^a}$ and $\boldsymbol{\tau^b}$ equivalent if they can be obtained one from the other through an opportune permutation of the indexes and we write $\boldsymbol{\tau^a} \sim \boldsymbol{\tau^b}$. We denote a class of equivalent outcomes as $\{\tilde{\boldsymbol{\tau}}\}$, where $\tilde{\boldsymbol{\tau}}$ is an ordered vector, i.e. a vector with increasing component ($\tilde{\tau}_1 \leq \tilde{\tau}_2 \leq \cdots \tilde{\tau}_n$). If $A$ is a set of ordered vectors, then $\{A\}$ denotes the union of the classes of equivalence of the vectors in $A$.

Performance perceived by a given station $i$ not only depends on the probability $\tau_i$ to access the channel, but also on the probability that no other station interferes on the same slot. Therefore, from the point of view of station $i$, the vector strategy $\boldsymbol{\tau}$ can be represented by the couple of values $(\tau_i, p_i)$, where $p_i = 1 - \prod_{j \neq i}(1 - \tau_j)$, the probability that at least another station transmits, summarizes the interactions with all the other mobile stations.

In presence of downlink traffic, we also assume that the AP contends for the channel as a legacy DCF station with saturated downlink traffic. Thus, the overall collision probability suffered by station $i$ results $1-(1-p_i)(1-\tau_{AP})$, where $\tau_{AP}$ is the channel access probability employed by the AP. Since the AP is a legacy station, its transmission probability is not chosen by the AP, but is function of the perceived collision probability $p_{AP}$, $\tau_{AP} = f(p_{AP})$, where $f()$ has been derived in ([7]):

$$\tau = f(p) = \begin{cases} \frac{2(1-p^{R+1})}{1-p^{R+1}+(1-p)\sum_{i=0}^{R} p^i W(i)} & 0 \leq p < 1 \\ \frac{2(R+1)}{1+\sum_{i=0}^{R} W(i)} & p = 1 \end{cases} \tag{1}$$

where $R$ is the retry limit employed in the network and $W(i)$ is the contention window at the $i_{th}$ retry stage (i.e. $W(i) = \min\{2^i CW_{min}, CW_{max}\}$). We can evaluate the AP collision probability as a function of the vector strategy $\boldsymbol{\tau}$ or as a function of a generic couple $(\tau_i, p_i)$:

$$p_{AP} = 1 - \prod_{i=1}^{n}(1 - \tau_i) = 1 - (1 - p_i)(1 - \tau_i)$$

### B. Station Utility

According to the slotted channel scale, the random access process can be described as a sequence of slots resulting in a successful transmission (when only one station accesses the channel), in a collision (when two or more stations access the channel), or in an idle slot (when no station accesses the channel). By observing that each slot boundary represents a regeneration instant [21] for the access process, the throughput of each station can be readily evaluated as the ratio between the average number of bits transmitted in each slot and the average duration of each slot [20].

Assuming that the AP equally shares the downlink throughput among the stations, we can express the uplink throughput $S^i_u$ and the downlink throughput $S^i_d$ for the $i$-th station as [20]:

$$S^i_u(\tau_i, p_i) = \frac{\tau_i(1 - p_i)(1 - \tau_{AP})P}{P_{idle}\sigma + [1 - P_{idle}]T} \tag{2}$$

$$S^i_d(\tau_i, p_i) = \frac{1}{n}\frac{\tau_{AP}(1 - p_{AP})P}{P_{idle}\sigma + [1 - P_{idle}]T} \tag{3}$$

where $P$ is the frame payload which is assumed to be fixed, $\sigma$ and $T$ are, respectively, the empty and the busy slot duration[1], and $P_{idle}$ is the probability that neither the stations, nor the AP transmit on the channel, i.e. $P_{idle} = (1 - p_{AP})(1 - \tau_{AP})$.

Since the downlink throughput is equal for all the stations, we can avoid the $i$ superscript in (3). We

[1]We are implicitly considering a basic access scheme, with EIFS=ACK_Timeout+DIFS, which corresponds to have a fixed busy slot duration in both the cases of successful transmission and collision.

define the utility function $J_i$ for the mobile station $i$ as:

$$J_i = \min\{S_u^i, kS_d\} \quad (4)$$

The rationale of this definition is the assumption that the station applications can require bandwidth on both directions. The coefficient $k \in [0, \infty)$ takes into account the desired ratio between the uplink and the downlink throughput. We assume that the application is the same for all the stations, thus using a fixed $k$ value for all the utility functions. When $k = 1$ all the stations require the same throughput in both directions. Note that $k = \infty$ corresponds to a *unidirectional traffic* case, in which stations are not interested in downlink throughput and their utility is simply given by the uplink throughput (as assumed in most previous literature). Figure 2 plots the utility of a given station $i$ for $k = 1$, in case of 802.11b physical (PHY) layer, $P = 1500$ bytes, a data rate equal to 11 Mbps, and an acknowledgment rate of 1 Mbps. In such a scenario, by including physical preambles, acknowledgment transmissions, MAC headers and interframe times, the $T$ duration is equal to 1667 $\mu s$. Different network conditions, summarized by different values of the $p_i$ probability, have been considered. The collision probability $p_i$ takes into account only the competing mobile stations, so that the actual collision probability is given by $1-(1-p_i)(1-\tau_{AP}(p_i, \tau_i))$. From the figure, it is evident that, for each $p_i$, the utility is maximized for a given best response value (e.g. about 0.01 for $p = 0.15$), which slightly decreases as $p_i$ grows.

We also consider the single variable functions $S_u^{hom}(\tau) = S_u(\tau, 1-(1-\tau)^{n-1})$ and $S_d^{hom}(\tau) = S_d(\tau, (1-\tau)^{n-1})$ representing, respectively, the uplink and downlink throughput perceived by each station in case of homogeneous outcomes $(\boldsymbol{\tau}|\tau_i = \tau, \forall i)$. Figure 3 plots the utility of a given station in case of homogeneous outcomes for $n = 2$ and $n = 10$, and for different $k$ values. In these curves $p_i = 1-(1-\tau)^{n-1}$ is not fixed, because the strategy changes are not unilateral. The optimal strategy, which maximizes the station utility, is function of both $n$ and $k$.

## C. Nash Equilibria

We are interested in characterizing Nash Equilibria (NE) of our game model where stations achieve a non-null utility. The inefficient equilibria in which all stations achieve an utility value equal to 0 can be easily found by observing that:

*Remark 2.1:* In general, station $i$ utility is a function of the whole set of strategies ($\boldsymbol{\tau}$), but it is constant and equal to 0 if a) $p_i = 1$, i.e. if at least one of the other players is transmitting with probability 1 ($\exists j \neq i \mid \tau_j = 1$), or if b) $\tau_i = 0$. The AP access probability $\tau_{AP}$ depends on $\tau_i$ and $p_i$ according to (1) and cannot be equal to 1 for standard contention window values.

*Proposition 2.1:* The vectors of strategies $\boldsymbol{\tau}$, such that $\exists\, i, j \in 1, 2, \cdots n \mid \tau_i = 1, \tau_j = 1$ are NE of the distributed access game in which all stations achieve an utility value equal to 0.

*Proof:* The result is an immediate consequence of Remark 2.1. If there are at least two stations transmitting with probability 1, then the channel is entirely wasted because of collisions and $S_u^l = S_d = 0, \forall l$. In these conditions, $J_l = 0\ \forall l$ and stations are not motivated in changing their strategies. The equilibrium strategies are then $\{(\boldsymbol{x}, 1, 1), \boldsymbol{x} \in [0,1]^{n-2}\}$. ■

The following remark will be useful for characterizing more efficient NE.

*Remark 2.2:* Consider a generic station $i$ and the collision probability $p_i \in (0, 1)$ suffered because of the other station strategies. By derivation, it can be easily proved that $S_d(\tau_i, p_i)$ is a monotonic decreasing function of $\tau_i$, starting from $S_d^i(0, p_i) > 0$, and that $S_u^i(\tau_i, p_i)$ is a monotonic increasing function of $\tau_i$, starting from $S_u^i(0, p_i) = 0$.

Let us consider now the *best response* strategy of a station $i$ and denote it by $\tau_i^{(br)}$. For $k = \infty$, the station utility function is equal only to $S_u^i(\tau_i, p_i)$. From Remarks 2.1 and 2.2, it results that the utility is maximized for $\tau_i^{(br)} = 1$ when $p_i < 1$, and it is constant to 0 when $p_i = 1$. For $k < \infty$, from Remark 2.2 we can state that the utility $J_i$ is maximized for $\tau_i^{(br)} \in (0, 1)$ such that $S_u^i(\tau_i^{(br)}, p_i) = kS_d^i(\tau_i^{(br)}, p_i)$. It follows that $\tau_i^{(br)}$ is the solution of the following implicit equation:

$$\tau_i^{(br)} = \frac{k\tau_{AP}}{n-(n-k)\tau_{AP}} = \frac{kf\left(1-(1-p_i)\left(1-\tau_i^{(br)}\right)\right)}{n-(n-k)f\left(1-(1-p_i)\left(1-\tau_i^{(br)}\right).\right)} \quad (5)$$

It can be shown that the previous equation has a single solution $\tau_i^*$ in the range $(0, 1)$, which can be numerically solved in a few fixed point iterations.

*Proposition 2.2:* In case of *unidirectional traffic* (i.e. $k = \infty$), the NE of the distributed access game with non-null utility values are all and only the vector of strategies $\boldsymbol{\tau}$, such that $\exists !\ i \in 1, 2, \cdots n \mid \tau_i = 1$.

*Proof:* Being $i$ the index of the single station transmitting with probability 1, it follows that $P_{idle} = 0$ and $p_i \neq 1$. Therefore, station $i$ achieves a non-null uplink rate equal to $S_u^i = P(1 - p_i)/T = J_i$. All the other stations achieve $S_u^j = 0$ for $j \neq i$ regardless of their strategy $\tau_j$. The vectors of strategies $\{(\boldsymbol{x}, 1), \boldsymbol{x} \in [0, 1)^{n-1}\}$ are then Nash Equilibria. Any other vector of strategies $\boldsymbol{\tau}$, such that $\tau_i \neq 1 \forall i$ cannot be a NE, since for $p_i \neq 1$, a generic station $i$ can improve its utility by increasing its strategy up to $\tau_i = 1$. ■

The vectors of strategies $\{(\boldsymbol{x}, 1), \boldsymbol{x} \in [0, 1)^{n-1}\}$ are not NE for $k < \infty$. In fact, being $p_{AP} = 1 - \prod_{h=1}^{n}(1 - \tau_h) = 1$, for these vectors of strategies the downlink throughput $S_d$ is equal to 0. Thus, the single station $i$ playing $\tau_i = 1$ is motivated to reduce its transmission probability to a value lower than 1, in order to increase its utility $J_i = \min\{S_u^i, S_d\}$.

*Proposition 2.3:* In case of *bidirectional traffic* (i.e. $k < \infty$), the homogeneous strategy vector $(\boldsymbol{\tau} | \tau_i = \tau^* = \frac{kf(1-(1-\tau^*)^n)}{n-(n-k)f(1-(1-\tau^*)^n)}, \forall i)$ is the only NE with non-null utility values of the distributed access game.

*Proof:* The strategy vector $(\tau^*, \tau^*, \cdots, \tau^*)$ is a NE as an immediate consequence of (5) for $p_i = 1 - (1 - \tau^*)^{n-1}$. In fact the equation can be read as a mutual best response. Since equation (5) has a single solution, there exists a unique homogeneous NE strategy. Since $(1 - p_i)(1 - \tau_i) = (1 - p_j)(1 - \tau_j), \forall i, j$, the right hand of the best response equation (5) is the same for all the stations. This excludes the existence of non-homogeneous Nash equilibria. ■

The parameter $\tau^*$, which characterizes the NE strategy with non-null utility only depends on the number of stations $n$ and desired ratio $k$ between uplink and downlink throughput. It is not affected either by the PHY layer parameters (such as backoff slot duration, interframe spaces, etc.) or by the frame length.

### D. Social utility

In this section, we try to identify desirable outcomes from a global point of view. A natural choice is to look at outcomes that maximize a global utility function, such as the minimum utility $J_S(\boldsymbol{\tau})$ perceived in the network: $\min_{i=1\cdots n} J_i$. This global utility is often referred to as social utility. The following remark will be useful for such a characterization.

*Remark 2.3:* The uplink throughput $S_u^{hom}(\tau)$ perceived in case of homogeneous outcomes is a non-monotonic function in $\tau$, with a single maximum value $S_u^{hom}(\tau_x)$, for $\tau_x \in (0, 1)$.

*Proposition 2.4:* The social utility is maximized for an homogeneous outcome $(\tau', \tau', \cdots, \tau')$ and such outcome is Pareto Optimal.

*Proof:* From the utility definition, we have that the minimum utility perceived in the network is given by $\min\{\min_{i=1,\cdots n} S_u^i, kS_d\}$. Therefore, the minimum utility can be due to the minimum uplink throughput or downlink throughput.

We observe that the minimum uplink throughput is maximized for a homogeneous outcome. In fact, if $\boldsymbol{\tau}$ is a non-homogeneous outcome in $A = \{\boldsymbol{x}, \boldsymbol{x} \in (0, 1)^n\}$, we can prove that the minimum uplink throughput cannot be maximized. Without loss of generality, we consider that $\boldsymbol{\tau}$ is an ordered vector, with $0 < \tau_1 < \tau_2 < \cdots < \tau_n < 1$ and $S_u^1(\tau_1, p_1) = \min_i S_u^i(\tau_i, p_i)$. Let $\tau_1'$ be a new strategy for station 1, such that $0 < \tau_1 < \tau_1' < \tau_2$. For the new outcome $\boldsymbol{\tau'} = (\tau_1', \tau_2, \cdots \tau_n)$, the minimum uplink throughput is still the throughput perceived by station 1. This throughput is higher than the previous one, since $S_u^i(\tau_i, p_i)$ is monotonic increasing in $\tau_i$. It follows that $S_u^1(\tau_1', p_1) = \min_i S_u^i(\tau_i, p_i') > S_u^1(\tau_1, p_1) = \min_i S_u^i(\tau_i, p_i)$. Therefore, $\boldsymbol{\tau}$ has to be homogeneous.

We also observe that when all the station utilities are limited by the downlink throughput, i.e. $S_u^i \geq kS_d \quad \forall i$, $S_d$ is maximized for an homogeneous outcome. In fact, a non-homogeneous outcome cannot maximize $S_d$, because it is always possible to increase the AP throughput, by slightly reducing a given channel access probability $\tau_j$ while maintaining $S_u^j \geq kS_d$ (i.e. while maintaining $\tau_j \geq \tau_j^{(br)}$).

We can conclude that the social utility is maximized for an homogeneous outcome. For homogeneous outcomes $(\boldsymbol{\tau} | \tau_i = \tau, \forall i)$, the social utility $J_S^{hom}(\tau)$ is a single variable function and can be expressed as $\min\{S_u^{hom}(\tau), kS_d^{hom}(\tau)\}$. Being $\tau^*$ the homogeneous NE strategy for which $S_u^{hom}(\tau) = kS_d^{hom}(\tau)$, and $\tau_x$ the homogeneous strategy defined in remark 2.3, $J_S^{hom}(\tau)$ has a single maximum in $\tau'$, with $\tau' = \tau^*$ for $\tau^* \leq \tau_x$, or $\tau' = \tau_x$ when $\tau_x < \tau^*$.

Therefore, $J_S^{hom}(\tau')$ is the maximum social utility achievable in the network.

Finally, we prove the Pareto optimality. We recall that a Pareto optimal outcome is one such that no-one could be made better off by changing the vector of strategies without making someone else worse off. Since the homogeneous strategy vector $\boldsymbol{\tau'}$ maximizes the minimum utility perceived in the network, any other vector of strategies different from $\boldsymbol{\tau'}$ will degrade the utility of the station perceiving the minimum performance. Let $\boldsymbol{\tau}$ be an outcome different from $\boldsymbol{\tau'}$, with at least a player better off than in $\boldsymbol{\tau'}$. This allocation is necessarily non-homogeneous because, for homogeneous outcomes, the maximum utility is reached in $\boldsymbol{\tau'}$. Since for non-homogeneous outcomes the minimum utility is lower than $J_S^{hom}(\tau')$, it exists a station $j$ (perceiving the minimum utility) whose payoff is lower at $\boldsymbol{\tau}$ than at $\boldsymbol{\tau'}$. It follows that $\boldsymbol{\tau'}$ is a Pareto outcome. ∎

Figure 3 shows some examples of $J^{hom}(\tau)$, where the maximum utility value $J_S^{hom}(\tau')$ is limited by the uplink throughput (i.e. $\tau^* > \tau_x$) or by the downlink one (i.e. $\tau^* \leq \tau_x$). Note that the intersection between the function $S_u^{hom}(\tau)$ and the function $kS_d^{hom}(\tau)$ depends on $k$. Figure 3 shows that the intersection strategy $\tau^*$, for which the utility function has an abrupt slope change, grows as the $k$ value increases. Let $k_x$ be the value of $k$ for which $\tau^* = \tau_x$. For $k \geq k_x$, $J_S^{hom}(\tau')$ is equal to $S_u^{hom}(\tau')$ and results maximized.

*Proposition 2.5:* If the solution $\tau^*$ of (5) for $p_i = 1 - (1 - \tau^*)^{n-1}$ is lower or equal to $\tau_x$, the NE $(\tau^*, \tau^*, \cdots \tau^*)$ is Pareto optimal.

*Proof:* For $\tau^* \leq \tau_x$, the NE is the homogeneous outcome that maximizes the minimum network utility. Therefore, from the previous preposition, it is also Pareto optimal. ∎

Figure 3 shows that the limit condition $\tau^* = \tau_x$ is approximately reached for $k_x = 20$ in case of $n = 2$, and for $k_x = 11$ in case of $n = 10$. For smaller $k$ values, the homogeneous NE $\boldsymbol{\tau^*}$ is Pareto optimal. For larger $k$ values, including the unidirectional traffic case $k = \infty$, the Pareto optimal outcome $\boldsymbol{\tau'}$ is not an equilibrium point and the NE $\boldsymbol{\tau^*}$ gives poor performance (i.e. performance much worse than $J_S^{hom}(\tau')$).

## III. Channel access mechanism design

Our previous considerations about equilibrium conditions and Pareto optimality have shown that the global performance of the distributed access scheme strongly depends on the desired ratio $k$ between downlink and uplink throughput. In fact, the station utility perceived in equilibrium conditions and the maximum social utility depend on $k$, i.e. on the intersection point between the curves $S_u^{hom}$ and $kS_d^{hom}$. However, such a result is based on the assumption that the Access Point behaves as a legacy station. In this section, we explore the possibility to use the Access Point for changing the $S_d$ or $S_u$ functions, in order to force desired equilibrium outcomes. Indeed, since the AP plays the role of gateway to external networks, it can also play the role of arbitrator for improving the global performance of its access network.

### A. Tuning of the AP channel access probability

A first solution for changing the uplink and downlink throughput curves is to use the AP channel access probability $\tau_{AP}$ as a configuration parameter. Then, $\tau_{AP}$ does not depend on $\boldsymbol{\tau}$ according to (1), but it is equal to a fixed value, which can be tuned by the AP. The best response (5) for all the stations is equal to

$$\tau^+ = \frac{k\tau_{AP}}{n - (n-k)\tau_{AP}} \tag{6}$$

and the NE in $(0, 1)^n$ becomes $(\tau^+, \tau^+, \cdots \tau^+)$.

Figures 4 and 5 show the effects of the $\tau_{AP}$ tuning on the utility perceived for homogeneous outcomes. We considered a scenario with 10 contending stations, a packet size of 1500 byte, and $k = 1$. Each labeled curve refers to a different $\tau_{AP}$ setting, as indicated in the legend. For each curve, the NE corresponds to the cuspid point. For comparison, the figures also plot a bold dashed curve for the case in which the AP behaves as a legacy station. Figure 4 has been obtained for an 802.11b PHY layer at the maximum transmission rate (namely, 11 Mbps). In this case, the utility perceived with a legacy AP at the NE is about the same perceived with a fixed $\tau_{AP} = 0.064$. However, when the packet transmission time is shorter, it is possible to improve the NE utility by opportunistically tuning a fixed $\tau_{AP}$ value. Figure 5 has been obtained for an 802.11n PHY layer at the maximum rate (namely, 600 Mbps). In this case, the difference between the NE utility perceived under a legacy AP and under a fixed $\tau_{AP} = 0.168$ is about $0.9$ Mbps.

Since each different $\tau_{AP}$ setting leads to a different homogeneous NE, given a desired NE it is possible to design the corresponding $\tau_{AP}$ value by inverting (6). We can express the utility $J^{NE}$ perceived at the NE as a function of the desired homogeneous NE $(\tau, \tau, \cdots \tau)$. By considering that at the NE the utility can be expressed as $S_u^{hom}$, (or equivalently as $kS_d^{hom}$), after some manipulations it results:

$$J^{NE}(\tau) = \frac{\tau(1-\tau)^n P}{T - (1-\tau)^{n+1}(T-\sigma) + \frac{n-k}{k}T\tau} \quad (7)$$

Figure 6 plots some examples of the NE utilities $J^{NE}$ achievable for $n = 10$, $P = 1500$ bytes, and different $k$ values, in the case of 802.11b PHY at 11 Mbps. By deriving (7), it can be shown that, for $k \neq 0$, the function $J^{NE}$ has a unique maximum in $\tau_o \in (0, 1)$. Such desired maximum can be obtained by setting a specific $\tau_{AP_o}$ value. Although an exact computation of $\tau_{AP_o}$ is in principle possible, by deriving $J^{NE}$ and inverting (6), an approximated optimal tuning can be derived as follows. In [7], it is shown that for a network with $N$ competing stations the optimal channel access probability is given by $\frac{1}{N\sqrt{T/2\sigma}}$. In our scenario, at the NE outcome, the AP downlink throughput corresponds to the aggregation of $n$ flows, whose bandwidth is a fraction $1/k$ of the uplink throughput perceived by each station. Therefore, we can consider the AP channel access probability as the aggregation of $n/k$ flows in a network with $n+n/k$ contending stations. It follows:

$$\hat{\tau}_{AP_o} = \frac{n}{k}\frac{1}{(n+n/k)\sqrt{T/2\sigma}} = \frac{n}{(n+kn)\sqrt{T/2\sigma}}, \quad (8)$$

which leads to the desired NE outcome

$$\hat{\tau}_o = \frac{k}{(kn+n)\sqrt{T/2\sigma} - (n-k)}.$$

Figure 6 visualizes the accuracy of the proposed approximation by plotting the points $(\hat{\tau}_o, J^{NE}(\hat{\tau}_o))$ (white boxes) for different $k$ values. The proposed approximation has an important practical implication. In fact, by simply specifying the application requirements $k$ and estimating the number of contending stations $n$ (as described in section IV), the access point can tune its channel access probability to $\hat{\tau}_{AP_o}$, thus forcing the network to the desired equilibrium point. We also report the NE utilities perceived under legacy AP (black boxes). Comparing white and black boxes, it is evident that in many cases the network performance under a legacy AP are suboptimal (i.e. the NE utility is lower than the maximum of the $J^{NE}$ curves.

The figure also plots the limit curve obtained for $k \to \infty$, which represents the system behavior when the application requirements tend to the unidirectional traffic case. In this case, the $J^{NE}$ expression given by 7 tends to the uplink throughput expression. However, this curve is practically unfeasible because for $k \to \infty$ $\tau_{AP}$ tends to zero for any desired NE, and cannot be used as a tuning parameter. In other words, although $\hat{\tau}_o$ tends to the finite value $1/(n\sqrt{T/2\sigma}+1)$ maximizing the social utility, the mechanism design cannot be practically performed. When $\tau_{AP} \neq 0$, such as in the legacy case, for $k \to \infty$ the best response of each station tends to 1 (as shown in the black boxes of figure 6).

### *B. ACK suppression*

The mechanism design described in the previous section is unfeasible when $k = \infty$. Moreover, when $k$ is very high (i.e. $\tau_{AP_o}$ tends to zero), stations need a long estimation time for correctly evaluate their best response. A solution for controlling the resource repartition in infrastructure networks with negligible (or zero) downlink throughput is a selective discard of the ACK transmissions at the AP side. Since the AP is the common receiver for all stations, suppressing the ACKs at the AP side corresponds to triggering ACK timeouts at the station side, which are interpreted as collisions. Therefore, the ACK dropping can act as a punishment strategy devised to limit the uplink throughput of too aggressive stations. We propose the following threshold scheme: if a generic station $i$ has an access probability $\tau_i$ higher than the a given value $\gamma$, the AP drops an ACK frame transmission with probability $\min\{\alpha(\tau_i - \gamma), 1\}$.

In this case, by considering $\tau_{AP} = 0$ (i.e. the unidirectional traffic case) or $\tau_{AP} \simeq 0$, the utility function $J_i$ of a given station $i$ can be expressed as:

$$J_i(\tau_i, p_i) = \begin{cases} \frac{\tau_i(1-p_i)}{P_{idle}\sigma+[1-P_{idle}]T} & 0 < \tau_i < \gamma \\ \frac{\tau_i(1-p_i)[1-\alpha(\tau_i-\gamma)]}{P_{idle}\sigma+[1-P_{idle}]T} & \gamma \leq \tau_i < \gamma + 1/\alpha \\ 0 & \gamma + 1/\alpha \leq \tau_i \leq 1 \end{cases} \quad (9)$$

where we recall that $P_{idle} = (1 - \tau_i)(1 - p_i)$. According to the previous expression, for $\tau_i \leq \gamma$ the utility function $J_i$ is an increasing function of $\tau_i$, while for $\tau_i \leq \gamma$ its slope depends on the $\alpha$ setting. By selecting an $\alpha$ value which corresponds to a negative derivative for $\gamma < \tau_i < \gamma + 1/\alpha$, the utility function is maximized for $\tau_i = \gamma$.

Figure 7 plots the station utility perceived in case of homogeneous outcomes, under the ACK suppression scheme, for n=10 and different $\alpha$ values. For $\alpha = 0$, the station utility is simply given by the uplink throughput. In this case, the utility is an increasing function of the channel access probability and the best response of each station is $\tau^{(br)} = 1$. For $\alpha > 0$, the utility function is maximized for $\tau^{(br)} < 1$. Such a maximum corresponds to $\gamma$ for large enough values of $\alpha$ (in figure, $\alpha = 80$).

We can then prove the following result.

*Proposition 3.1:* The outcome $(\boldsymbol{\tau'}|\tau_i = \tau', \forall i)$ is a Pareto optimal Nash equilibrium of the game, when the ACK suppression scheme indicated above is implemented with

$$\begin{aligned} \gamma &= \tau', \\ \alpha &\geq \frac{1}{\tau'(1 + \tau'(-1 + \frac{T}{T-(T-s)(1-\tau')^{n-1}}))}. \end{aligned} \quad (10)$$

*Proof:* First we observe that $\boldsymbol{\tau'}$ is a NE. In fact, whatever player we consider, say it player $i$, Remark 2.1 guarantees that for $\tau_i < \tau'$ $J_i$ decreases as $\tau_i$ decreases. For $\tau' < \tau_i < \tau' + 1/\alpha$ inequality (10) guarantees that $J_i$ decreases as $\tau_i$ increases until it does not reach the value 0. For $\tau > \tau' + 1/\alpha$, the punishment strategy implies $J_i = 0$. Then deviating from $\boldsymbol{\tau'}$ is not convenient for player $i$.

Second, $\boldsymbol{\tau'}$ is the unique point of maximum for the social utility $J_S^{hom}(\tau)$. In fact, the social utility is given by the minimum utility perceived in the network, which is given by $S_u^{hom}(\tau)$ in the absence of punishment. The punishment strategy can only decrease the social utility for $\boldsymbol{\tau} \in [0,1]^n - [0,\tau']^n$ and keeps it unchanged otherwise. Therefore, the preposition 3.1 holds also under the punishment strategy and the outcome $\boldsymbol{\tau'}$ is Pareto optimal. ■

The utility $J^{NE}$ perceived at the NE can be simply expressed as the uplink throughput perceived in case of homogeneous outcome $(\boldsymbol{\tau}|\tau_i = \tau\forall i)$:

$$J^{NE}(\tau) = \frac{\tau(1-\tau)^n P}{(1-\tau)^n\sigma + (1-(1-\tau)^n)T} \quad (11)$$

In [7] it is shown that such a function has a unique maximum for $\tau \in (0, 1)$, that can be approximated as:

$$\tau_o = \frac{1}{n\sqrt{T/2\sigma}} \quad (12)$$

Therefore, by estimating the number of contending stations $n$, each AP can implement (as described in section IV) an ACK suppression scheme that forces the system to work on the NE $\boldsymbol{\tau_o}$ maximizing the network throughput.

## IV. Game-based MAC scheme: Implementation and Evaluation

On the basis of the results discussed in the previous sections, we propose some simple DCF extensions devised to enable each contending station to dynamically tune its channel access probability according to a best response strategy. For this purpose, each station needs two estimators for probing the uplink and downlink load conditions. In fact, station best response depends not only on the application requirements (by means of $k$), but also on the uplink load (by means of $n$) and downlink load (by means of $\tau_{AP}$).

We consider both the case in which the AP behaves as a legacy station, and the case in which the AP acts as a game designer, for forcing desired equilibrium conditions. In this second case, we assume that also the AP is able to estimate the uplink load (by means of $n$) and the channel access probability $\tau_i$ employed by each station. These estimates are then used for opportunistically tuning the AP channel access probability or the ACK suppression scheme (by means of $\gamma$ and $\alpha$).

### A. Network load estimators

In contention-based networks, network load can be simply related to channel observations. Considering the slotted channel model due to saturation conditions, a channel observation corresponds to the channel outcome observed into a given slot. Such outcome is given by an idle slot when no station transmits, by a successful slot when a single station transmits, by a collision slot when two or more stations transmit simultaneously. In order to perform run-time estimators, the channel observations can be grouped in regular observation intervals at which new measurement samples are available. We express the measurement intervals in terms of an integer number $B$ of channel slots. Since slot size is uneven (because successful and collisions slots last for a $T$ time, while idle slots last only for $\sigma$), the actual time required for a new measurement sample is not fixed.

For measuring the number of stations actually contending on the network, we propose to count the number of different transmitters observed during the measurement interval [6]. Each transmitter can be identified by means of its MAC address. Obviously, the monitoring station can identify the transmitter address only when the packet is received correctly. Thus, within $B$ observation slots, the number of successful packets is not fixed. Let $n^m(t)$ be the uplink load measurement performed during the $t$-th measurement interval by a given station. The estimation $\hat{n}$ of the number of contending stations is then performed as:

$$\hat{n}(t) = \delta\hat{n}(t-1) + (1-\delta)n^m(t) \qquad (13)$$

where $\delta$ is the filter memory.

The filter performance is critically affected by the time window $B$. Indeed, $n^m$ is a cumulative parameter, which does not allow to identify all the different transmitters summed up to $n^m$. Therefore, the filter memory is not able to correct a partial counting of the actual number of contending stations during $B$. In other words, $B$ has to be carefully tuned, in order to guarantee a reasonable probability to catch all the contending stations in each measurement interval. Consider for example the case in which $n = 4$ and $B$ is set to a value able to include only two successful transmissions. Assuming that stations 1 and 3 transmit during the first measurement interval, stations 2 and 4 during the second one, and so on, the filter will consider $n^m(0) = n^m(1) = \cdots n^m(t) = 2$, thus providing a wrong estimate of $n$. A solution can be an automatic tuning of $B$ according to the comparison of measurements performed during different observations intervals (e.g. $B$ and $2B$). In our previous example, by counting $n^m$ during $B$ and during $2B$, we find $n^m(B,0) = 2, n^m(2B,0) = 4$. Since the measurement performed during $2B$ is higher than the one performed during $B$, the station understands that $B$ has to be increased. Due to space constraints, we omit the details of the automatic tuning of $B$.

In order to measure the channel access probability employed by the AP, the monitoring station has to count the number $tx_{AP}$ of successful transmissions performed by the AP during $B$. Given that there is no way of understanding which station has transmitted in a collision slot, the station has also to count the total number of collisions $C$ for measuring the $\tau^m_{AP}(t)$ parameter in the $t$-th time interval as $\frac{tx_{AP}}{B-C}$. The estimation $\hat{\tau}_{AP}$ is then performed as:

$$\hat{\tau}_{AP}(t) = \beta\hat{\tau}_{AP}(t-1) + (1-\beta)\tau^m_{AP}(t) \qquad (14)$$

where $\beta$ is the filter memory. Note that the $B$ setting is not critical for this estimator.

### *B. Best response performance under legacy AP*

As discussed in section II, in case of unidirectional traffic, the best response strategy leads to very poor throughput performance under legacy AP. Therefore, we analyze the case $k = \infty$ only in the next subsection, when the AP acts as a game designer.

For $k < \infty$, a generic station $i$ may implement a best response strategy, on the basis of the previous estimators and (5), by setting its channel access probability to:

$$\tau^{(br)}(t+1) = \frac{\hat{\tau}_{AP}(t)}{\hat{n}(t) - (\hat{n}(t) - k)\hat{\tau}_{AP}(t).} \qquad (15)$$

In order to evaluate the effectiveness of this scheme (approximating the performance of an ideal best response in which all stations exactly know the network load), we run some simulations. We extended the custom-made C++ simulation platform used in [7]. We considered an 802.11g physical rate, with the data rate set to 6Mbps. The contention windows used by the AP have been set to the legacy values $CW_{min} = 16$ and $CW_{max} = 1024$. All the simulation results have been obtained by averaging 10 different simulation experiments lasting 10s, leading to a confidence interval lower than 3%. Unless otherwise specified, the measurement interval $B$ has been set to 500 channel slots (which averagely correspond to 300ms). Figure 8 compares the behavior of our scheme with standard DCF.

Each point refers to a network scenario in which $n$ stations (indicated in the $x$ axis), with an application requirement $k$, compete on the channel with a legacy AP. The aggregated uplink throughput (i.e. the sum of the throughput perceived by all the mobile stations) and $k$ times the aggregated downlink throughput, (i.e. the AP throughput) are indicated by the $y$ axis, respectively by white and black points. From the figure, it is evident that, as the number of contending stations increases, standard DCF gives very poor performance to the downlink throughput. Conversely, for $k = 1$ our scheme is able to equalize uplink and downlink throughput for each $n$, and even in congested network conditions. Moreover, it is also able to maintain the overall network throughput (i.e. the sum of the aggregated uplink and downlink throughput) almost independent on the network load. For example, for $n = 20$ the sum of the uplink and downlink throughout is about 3.8 Mbps for standard DCF and about 5 Mbps for our scheme. The figure also proves our scheme effectiveness for different application requirements (i.e. $k = 0.5$). The figure clearly visualizes that $\sum_i S_u^i = knS_d$ as expected. We also considered our scheme performance in time-varying load conditions, by running several simulation experiments in which the number of contending stations dynamically changes during the simulation time. Figure 9 shows a simulation example lasting $300$ seconds, in which we start with $5$ contending stations, add $5$ more stations after $100$ seconds and switch $3$ stations off at $200$ seconds. In the figure we plot the throughput perceived by a given reference station (namely, station labeled as station $1$) and the aggregated throughput perceived by the AP. The figure also plots the number of active stations estimated by station 1 on the basis of the load filter defined in (13), for $\delta = 0.7$ and $B = 500$ slots. We can draw some interesting observations. First, the average AP throughput is basically independent on the number of stations contending in the network. This behavior is very different from the standard DCF behavior, according to which the AP behaves as a normal station. Second, the station throughput is approximately equal to the desired ratio, i.e. to $1/n$ the downlink throughput.

Note also that our scheme is different from a classical prioritization scheme, such as the schemes defined in the EDCA extensions. Indeed, by giving lower contention windows to the AP (i.e. an higher EDCA priority class to the AP), it is not possible to perform a desired resource repartition between uplink and downlink which is also load independent. Finally, the load estimator gives quite stable results, because the number of different stations listened in each temporal window $B$ is always equal to the total number of contending stations. We can see that only in the interval between 100 and 200 seconds, during which we have a number of contending stations equal to 10, the estimator gives a number of contending stations lower than the actual one for two times. In fact, since we are monitoring a fixed number of slots, as the number of stations in the network grows, it is more likely that in some intervals some stations do not succeed in transmitting on the channel.

### C. *Best response performance under AP mechanism design*

In order to enable the AP to play the role of game designer, it is necessary to equip this central node with different estimation functions. In case of bidirectional traffic, the AP has to estimate the number of competing stations $n$ for tuning its channel access probability to the optimal value given by (8). For this purpose, the same estimator defined in (IV-A) can be used at the AP side. We assume that the application requirement $k$ does not need to be estimated, because it is known a priori (e.g. by means of signaling messages). In turns, each contending station runs an estimator of the AP channel access probability in order to perform a best response strategy adjustment according to (6).

Figure 10 shows an example of uplink and downlink throughput repartition when the AP implements a mechanism design scheme. The simulation refers to an experiment lasting 110 seconds, with 10 competing stations, a packet payload of 1500 byte, an 802.11 PHY layer at 11 Mbps, and $k = 0.5$. For comparison, the figure also plots the throughput repartition perceived under a legacy AP. The figure shows that the mechanism design implementation is effective in improving the downlink and uplink throughput performance. Such an improvement can be obtained despite of the simplicity of the employed load estimators.

When $k \to \infty$, the implementation of the ACK suppression scheme requires that the AP evaluates: i) the $\gamma$ threshold, which depends on the estimate of $n$; ii) the $\alpha$ coefficient, which is simply related to $\gamma$ and to the PHY parameters $T$ and $\sigma$; iii) the per-station channel access probability $\tau_i$, $i = 1, 2, \cdots n$, which require $n$ estimators similar to the one used by the stations for evaluating $\tau_{AP}$. During the observation interval $B$, the AP has to separately count the successful transmissions $tx_i$ performed by each station $i$, and the number of collisions $C$, for measuring $\tau_i^m$ as $tx_i/(B-C)$. These measurements can be filtered with the usual auto-regressive filter. The implementation of the ACK suppression scheme at the AP side has important implications for preventing users and card manufacturers from using non-standard contention window values. As proved in [3], currently there is an impressive proliferation of cheating cards, i.e. cards which implement lower contention windows for taking advantage during the contention with other cards.

Figure 11 shows a simulation example (reproducing one of the realistic scenarios documented in [3]), in which a cheater card with a contention window equal to 8 compete with one legacy card. The figure refers to a simulation experiment lasting 105 seconds, after a transient phase of 10 seconds. Despite of the temporal fluctuations, it is evident that the cheating card obtains a throughput (dashed line) higher than two times the throughput (bold line) perceived by the legacy card. Figure 11 also plots the throughput performance of the two stations when the AP implement the ACK suppression scheme. In this case, the cheater is no longer motivated to use a channel access probability higher than the contending station, since its throughput is maximized tuning the channel access probability to the threshold value $1/2\sqrt{T/2\sigma}$ (i.e. implementing a best response strategy). The figure shows the throughput performance of the two stations implementing the best response (bold and dashed lines labeled as best response), and the throughput degradation perceived by the cheater station by still using a contention window equal to 8. The ACK suppression scheme works properly, even if the AP relies on the channel access probability estimators, rather than on the actual values.

In order to asses the effectiveness of the ACK suppression implementation as the number of competing stations grows, figure 12 compares the aggregated network throughput of our scheme with the standard DCF one, for different $n$ values. Each point refers to a network scenario in which $n$ stations (indicated in the $x$ axis) are aware of the ACK suppression risk and employ a consequent best response strategy. Although the variance of the $\tau_i$ estimators could imply that in some intervals such a probability passes the threshold value (i.e. there is a non null probability of unnecessary ACK dropping), the figure shows that the aggregated throughput is almost constant regardless of the number of competing stations. This behavior is very different from standard DCF, whose efficiency depends on the number of contending stations and degrades for high load conditions. Therefore, our scheme is able not only to discourage cheating card behaviors, but also to optimize the global network performance.

## V. Conclusions

The proliferation of MAC-level programmable WiFi cards can potentially create serious coexistence problems, since some stations can implement greedy access policies for increasing their bandwidth share at the expenses of compliant users. For this reason, we proposed a game-theoretic analysis of persistent access schemes for WiFi infrastructure networks, in order to characterize equilibria conditions and to design disincentive mechanisms for inefficient behaviors. We proved that, when stations are interested in both uploading and downloading traffic, an homogeneous Nash Equilibrium arises, where all the stations reach the same non-null utility. Moreover, we also explored the utilization of the Access Point as an arbitrator for improving the global network performance. Specifically, we proposed two different solutions. When the required download traffic is different from zero, the AP can simply tune its channel access probability for controlling the station best response. When the downlink traffic is zero (or negligible), the AP can selectively discard the acknowledgments of too greedy stations.

We proposed some extensions to standard DCF, in order to i) estimate the network status, and ii) emulate an access scheme based on best response strategies and AP mechanism design. We proved the effectiveness of our solutions for controlling the resource sharing in WiFi networks in various network scenarios. We are currently investigating on the prototyping of our solutions in actual WiFi cards and APs. While the estimate and best response modules can be simply implemented at the driver level, the ACK dropping scheme requires a hardware/firmware update.

# References

[1] IEEE Standard 802.11 - 1999; Wireless LAN Medium Access Control (MAC) and Physical Layer (PHY) Specifications; November 1999.

[2] Wi-Fi Alliance, www.wi-fi.org

[3] G. Bianchi, A. Di Stefano, C. Giaconia, L. Scalia, G. Terrazzino, I. Tinnirello, "Experimental assessment of the backoff behavior of commercial IEEE 802.11b network cards", *Proc. of IEEE Infocom 2007*, May 2007, Anchorage, pp. 1181-1189.

[4] S.W. Kim, B.S. Kim, Y. Fang, "Downlink and uplink resource allocation in IEEE 802.11 wireless LANs", *Proc. of IEEE Conf. on Vehicular Technology*, Jan. 2005, Dallas, vol. 54, pp. 320-327.

[5] IEEE 802.11e Supplement to Part 11: Wireless Medium Access Control (MAC) and Physical Layer Specification: Medium Access Control (MAC) Enhancements for Quality of Service (QoS)", October 2005.

[6] L. Zhao, L. Cong, H. Zhang, W. Ding, J. Zhang, "Game-Theoretic EDCA in IEEE 802.11e WLANs", *Proc. of IEEE VTC 2008*, Sept. 2008, pp. 1-5.

[7] G. Bianchi, "Performance Analysis of the IEEE 802.11 Distributed Coordination Function", *IEEE Journal of Selected Areas in Communication*, vol. 18, no. 3, pp.535-547, March 2000.

[8] L. Giarré, I. Tinnirello, and G. Neglia, "Medium Access in WiFi Networks: Strategies of Selfish Nodes", *IEEE Signal Processing Magazine (SPM)*, vol. 26, issue 5, pp. 124-127, 2009.

[9] J. Konorski, "A Game-Theoretic study of CSMA/CA under a backoff attack", *IEEE/ACM Trans. on Networking*, vol. 14, issue 6, pp. 1167-1178, 2006.

[10] M. Cagalj, S. Ganeriwal, I. Aad, J.P. Hubaux, "On selfish behavior in CSMA/CA networks", *Proc. of IEEE Infocom*, March 2005, Miami, vol. 4, pp. 2513-2524.

[11] L. Chen, S.H. Low, J. Doyle, "Contention Control: A Game-Theoretic Approach", *Proc. of IEEE Conf. On Decision and Control*, Dec. 2007, New Orleans, pp. 3428-3434.

[12] T. Cui, L. Chen, S. Low, "A Game-Theoretic framework for medium access control", *IEEE JSAC*, vol. 26, issue 7, pp. 1116-1127, 2008.

[13] G. Zhang, H. Zhang, "Modeling IEEE 802.11 DCF in wireless LANs as a dynamic game with incompletely information", *Proc. of Conf. on Wireless, Mobile and Multimedia Networks*, Jan. 2008, Mumbai, pp. 215-218.

[14] L. Chen, J. Leneutre, "Selfishness, not always a nightmare: modeling selfish MAC behaviors in wireless mobile ad-hoc networks", *Proc. of IEEE ICDCS 2007*, June 2007, pp. 16-23.

[15] D.K. Sanyal, M. Chattopadhyay, S. Chattopadhyay, "Improved performance with novel utility functions in a game-theoretic model of medium access control in wireless networks", *Proc. of IEEE TENCON 2008*, Nov. 2008, pp. 1-6.

[16] I. Tinnirello, L. Giarré and G. Neglia, "The Role of the Access Point in Wi-Fi Networks with Selfish Nodes", *Proc. of IEEE GameNets*, Instanbul, pp. 632-637, 2009.

[17] L. Giarré, G. Neglia and I. Tinnirello "Resource sharing optimality in WiFi infrastructure networks", *IEEE Proc. of CDC*, Shangai, 2009.

[18] G. Bianchi, I. Tinnirello, "A Kalman filter estimation of the number of competing terminals in an IEEE 802.11 network", *Proc. of IEEE Infocom*, April 2003, San Francisco, vol.2, pp. 844-852

[19] L. Giarré, G. Neglia and I.Tinnirello, "Performance analysis of selfish access strategy on WiFi infrastructure networks", *IEEE Proc. of Globecom*, Hawaii, 2009.

[20] G. Bianchi, I. Tinnirello "Remarks on IEEE 802.11 Performance Analysis", *IEEE Communication Letters*, Vol. 9, no. 8, August 2005.

[21] D.R. Cox, "Renewal Theory", Methuen and Co., 1970.

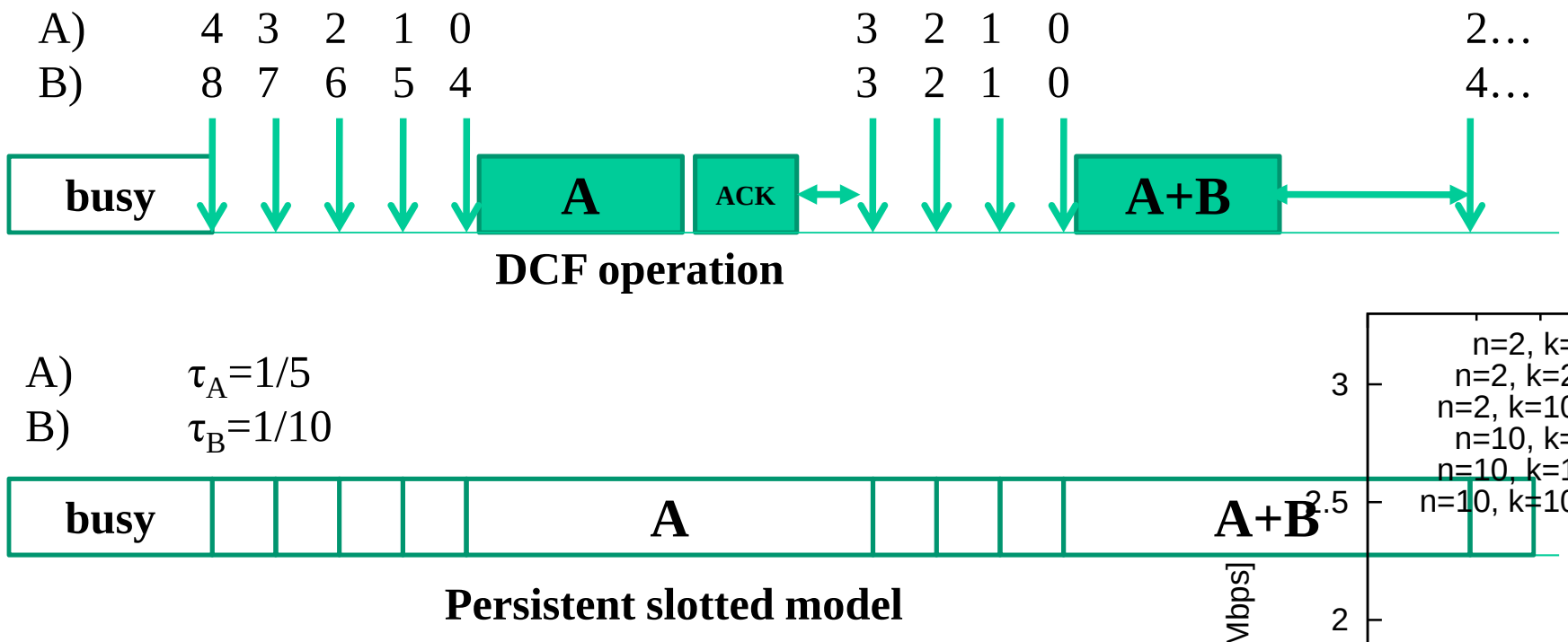

Fig. 1. An example of DCF operation and equivalent slotted model.

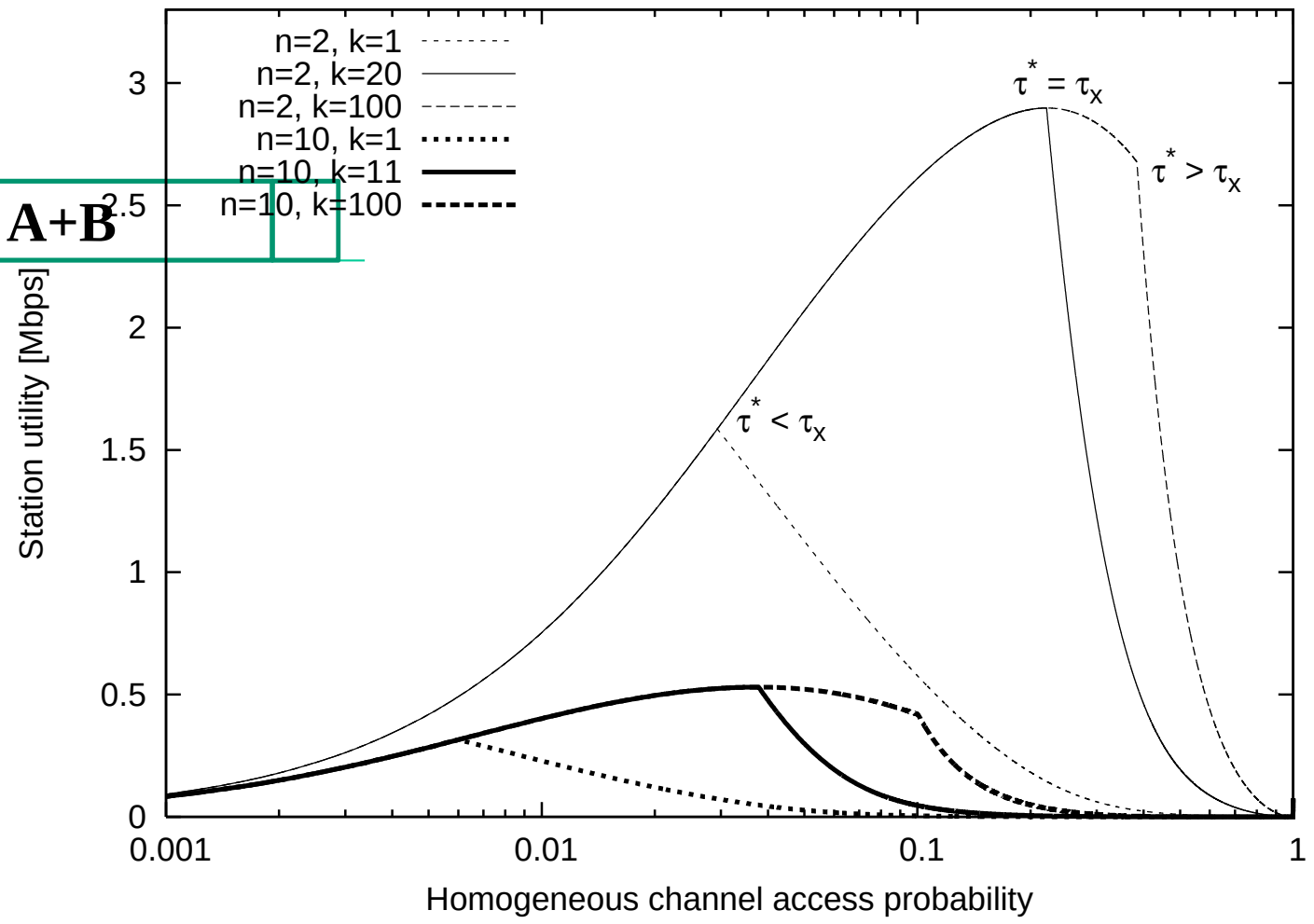

Fig. 3. Station utility in case of homogeneous access probability employed by all the stations and different $k$ values.

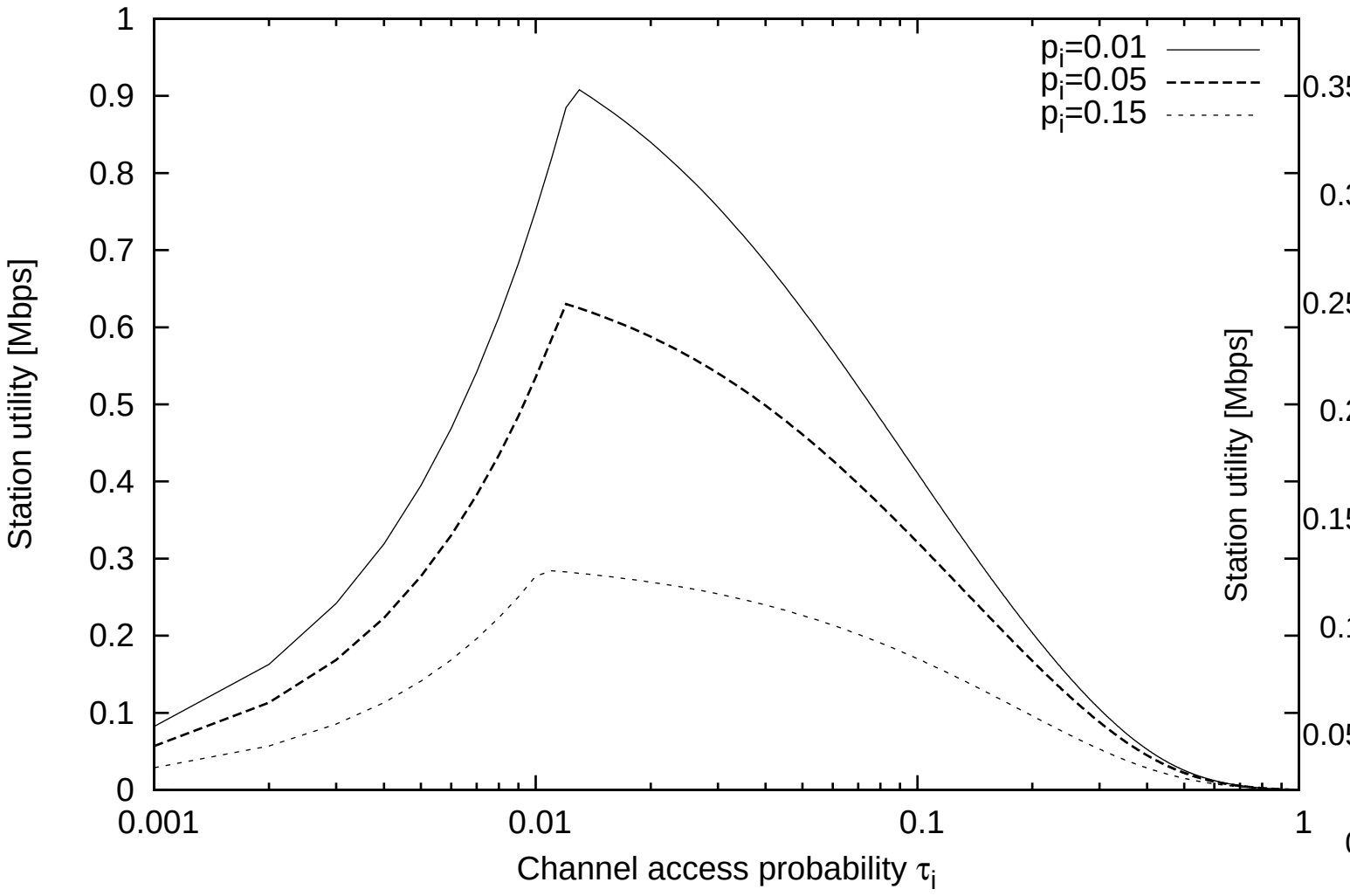

Fig. 2. Utility of a given station $i$, for different $p_i$ values, as a function of the strategy $\tau_i$ $(k = 1)$.

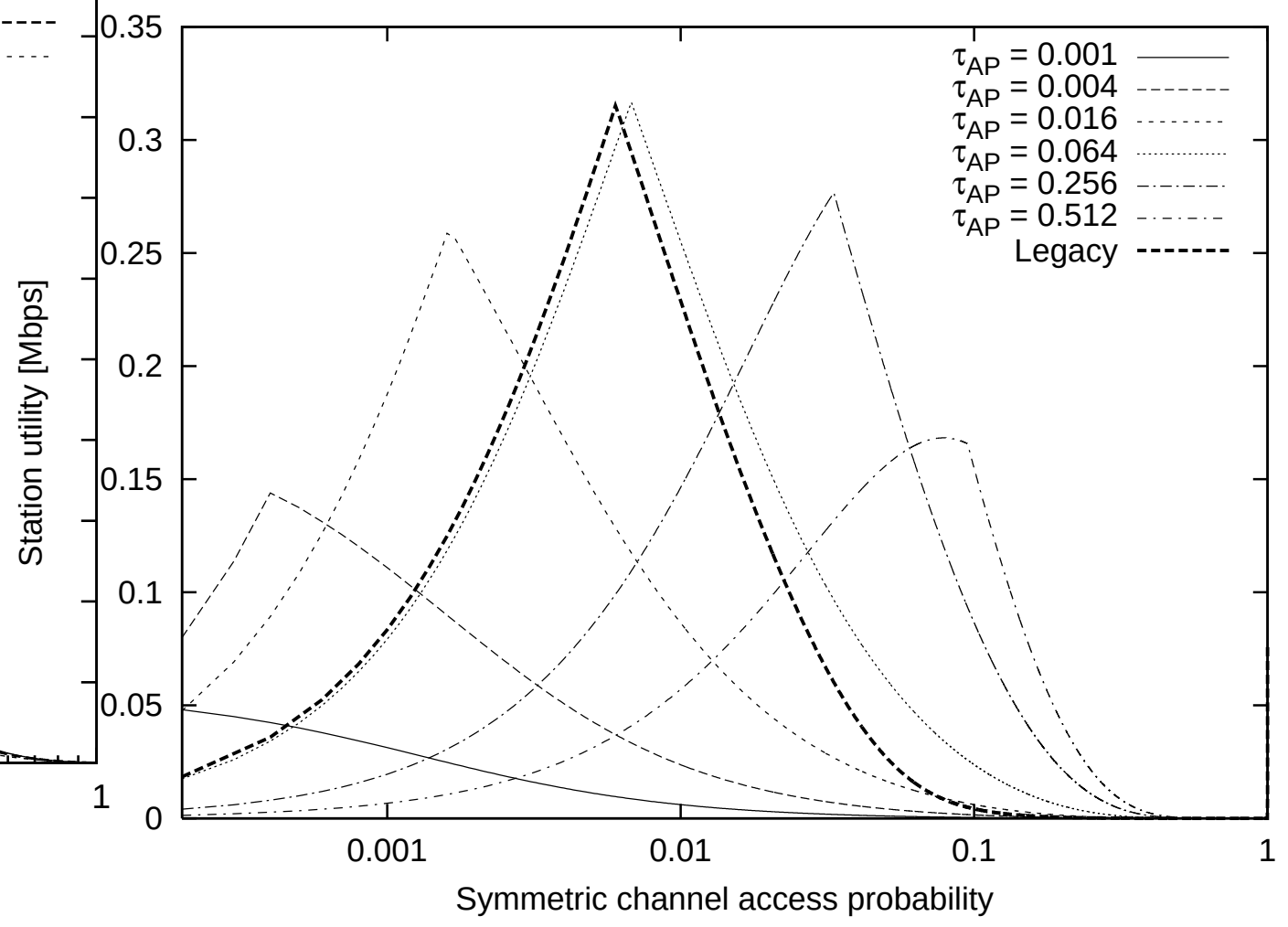

Fig. 4. Station utility at 11 Mbps in case of homogeneous access probability, n=10, k=1, and different $\tau_{AP}$ values.

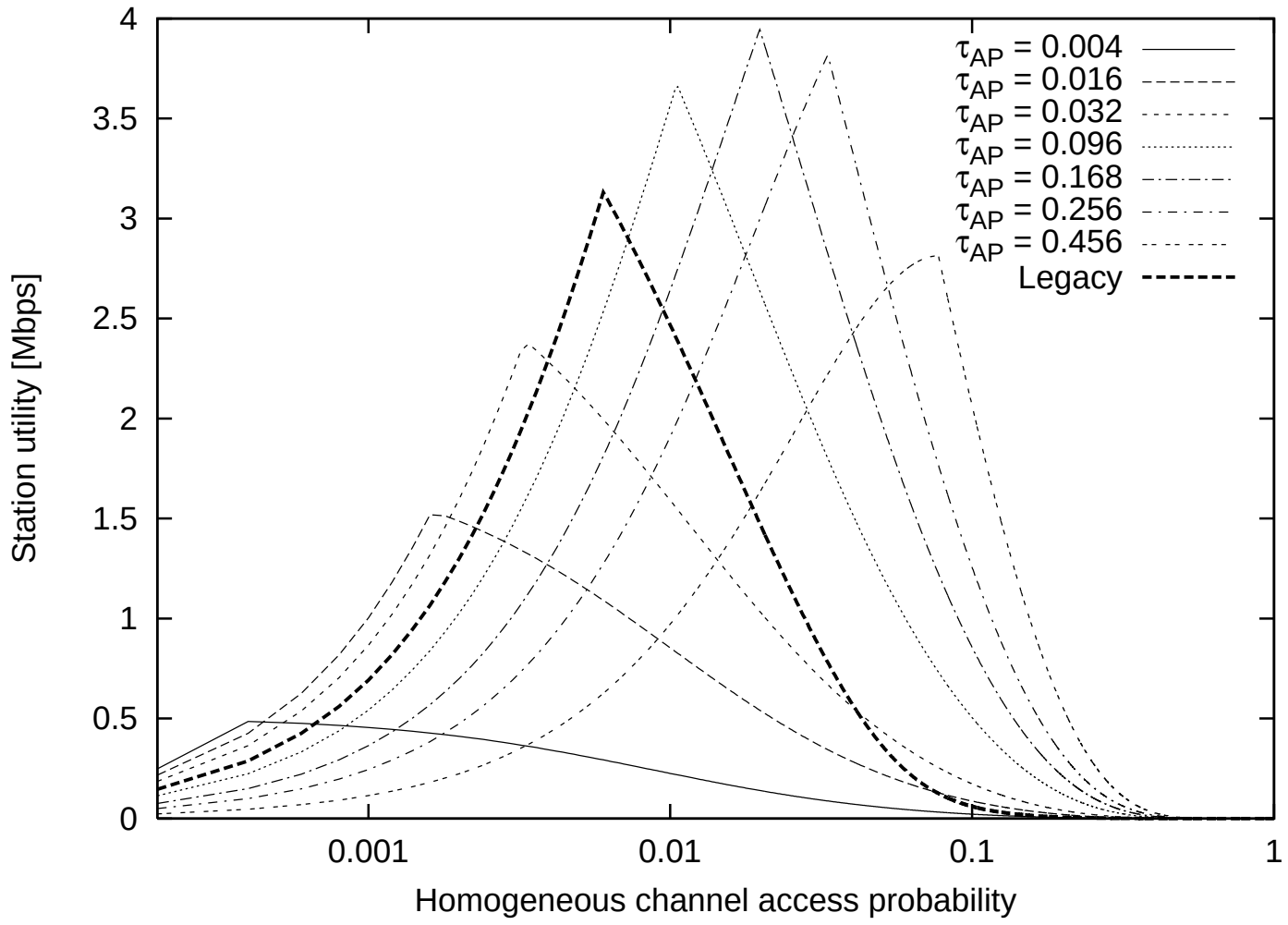

Fig. 5. Station utility at 600 Mbps in case of homogeneous access probability, n=10, k=1, and different $\tau_{AP}$ values.

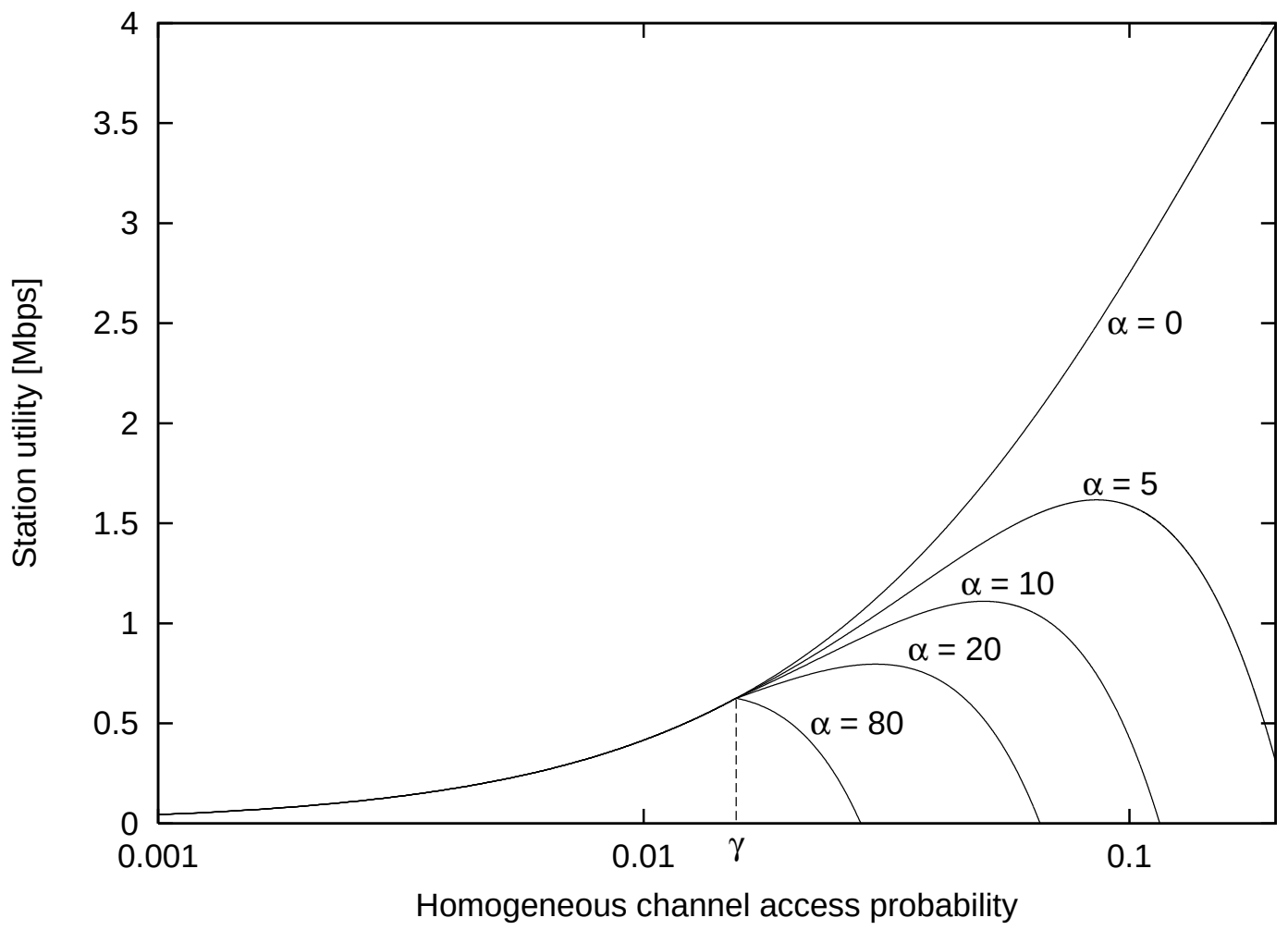

Fig. 7. Station utility in case of ACK suppression, for n=10, k=∞ and different $\alpha$ values.

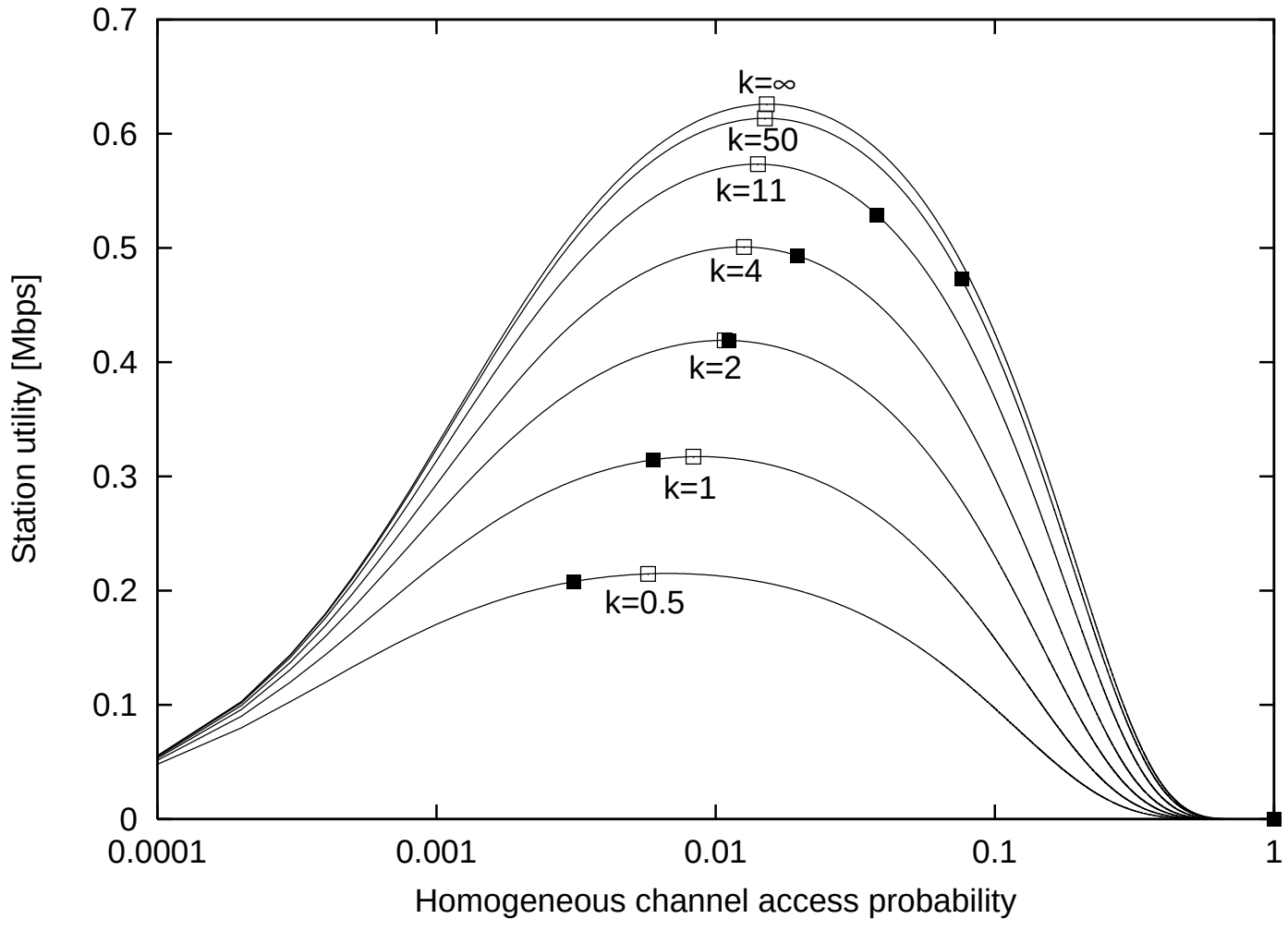

Fig. 6. Station utility at NE as a function of the desired NE outcome, for n=10 and different k values. Maximum station utility under mechanism design (white boxes), station utility at NE under legacy AP (black boxes).

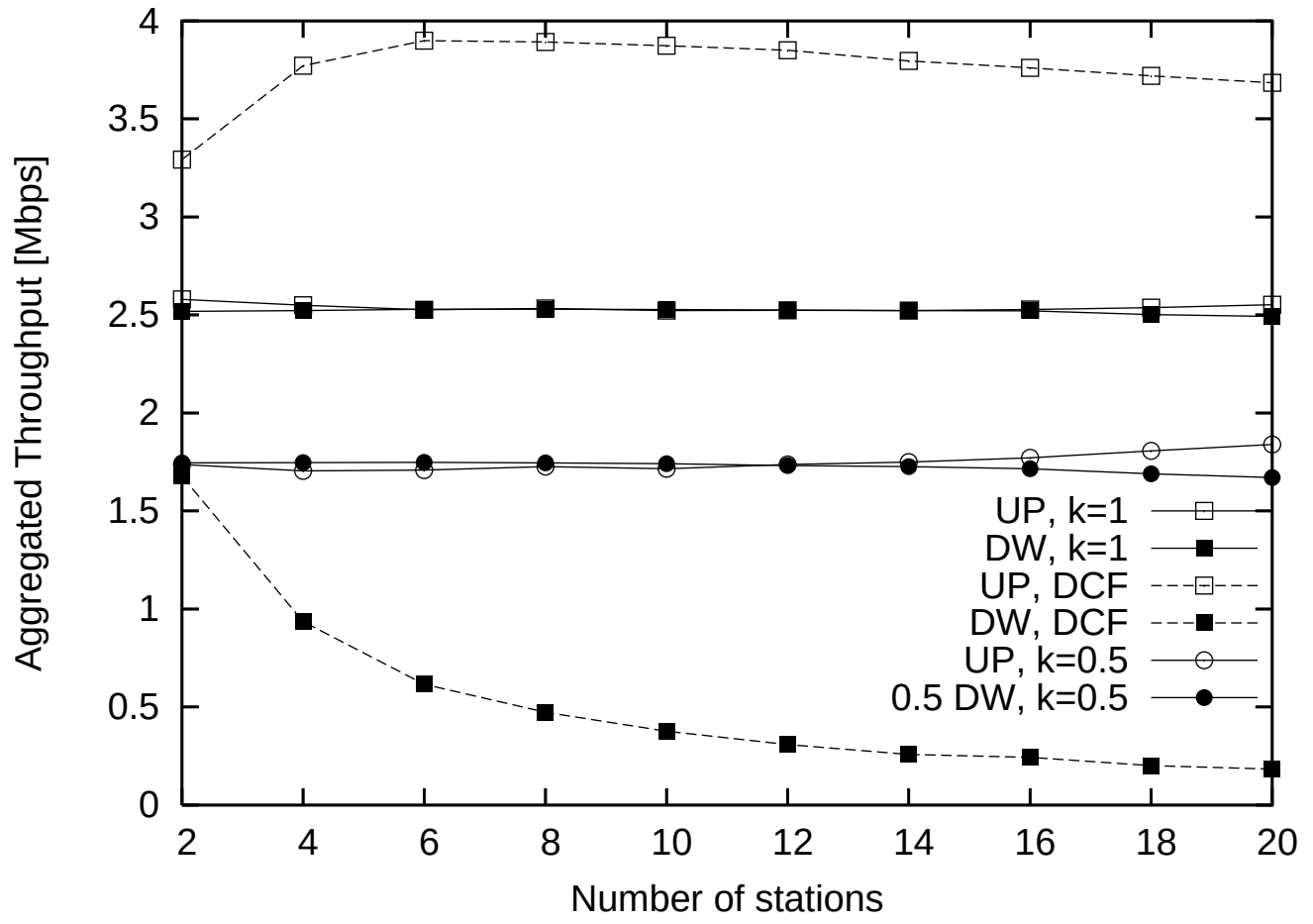

Fig. 8. Aggregated throughput for various number of nodes. Comparison of our scheme ($k = 1$, $k = 0.5$) with standard DCF.

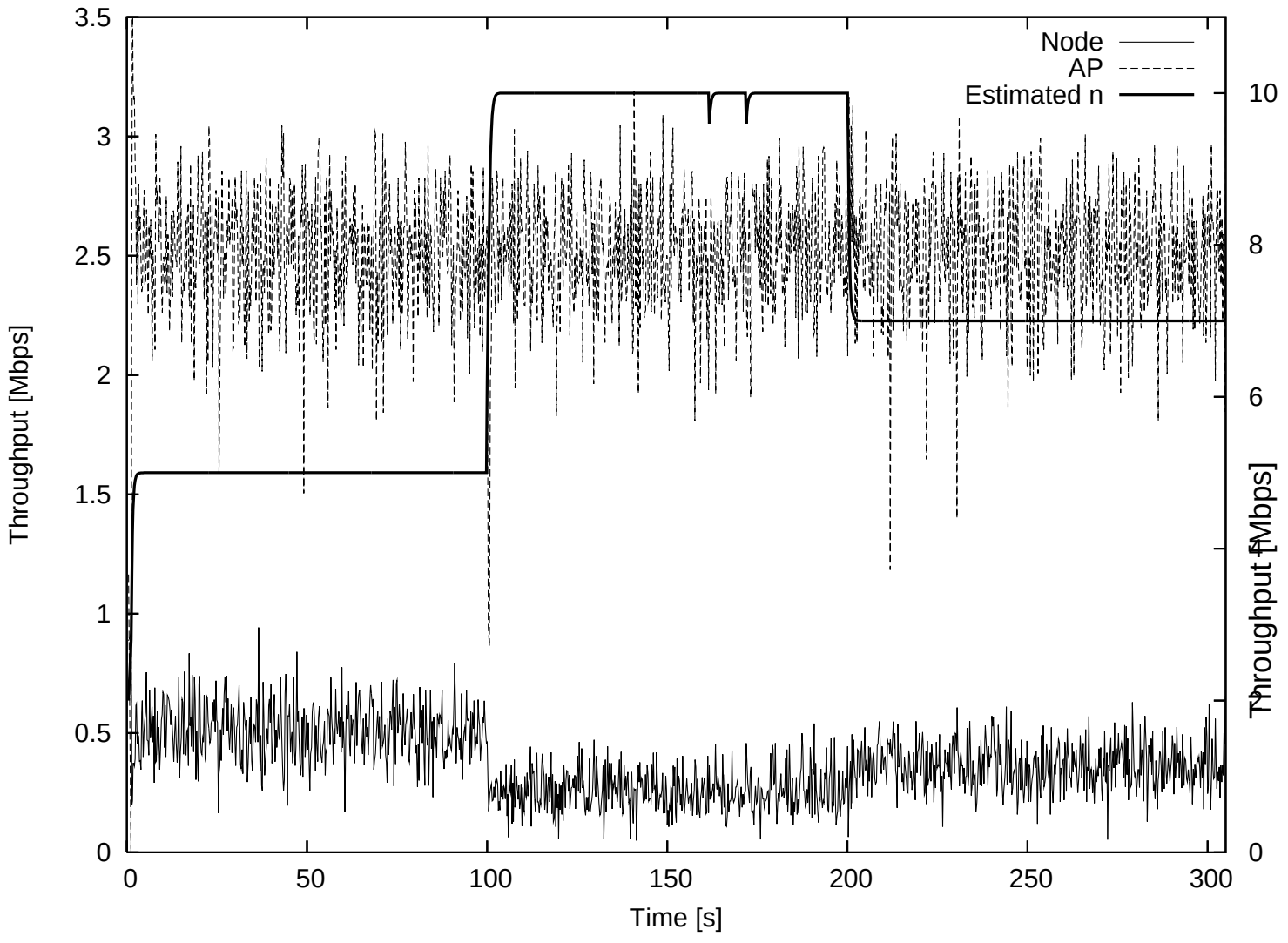

Fig. 9. Effects of best response strategies ($k = 1$) on AP aggregated throughput and a given node throughput in dynamic load conditions.

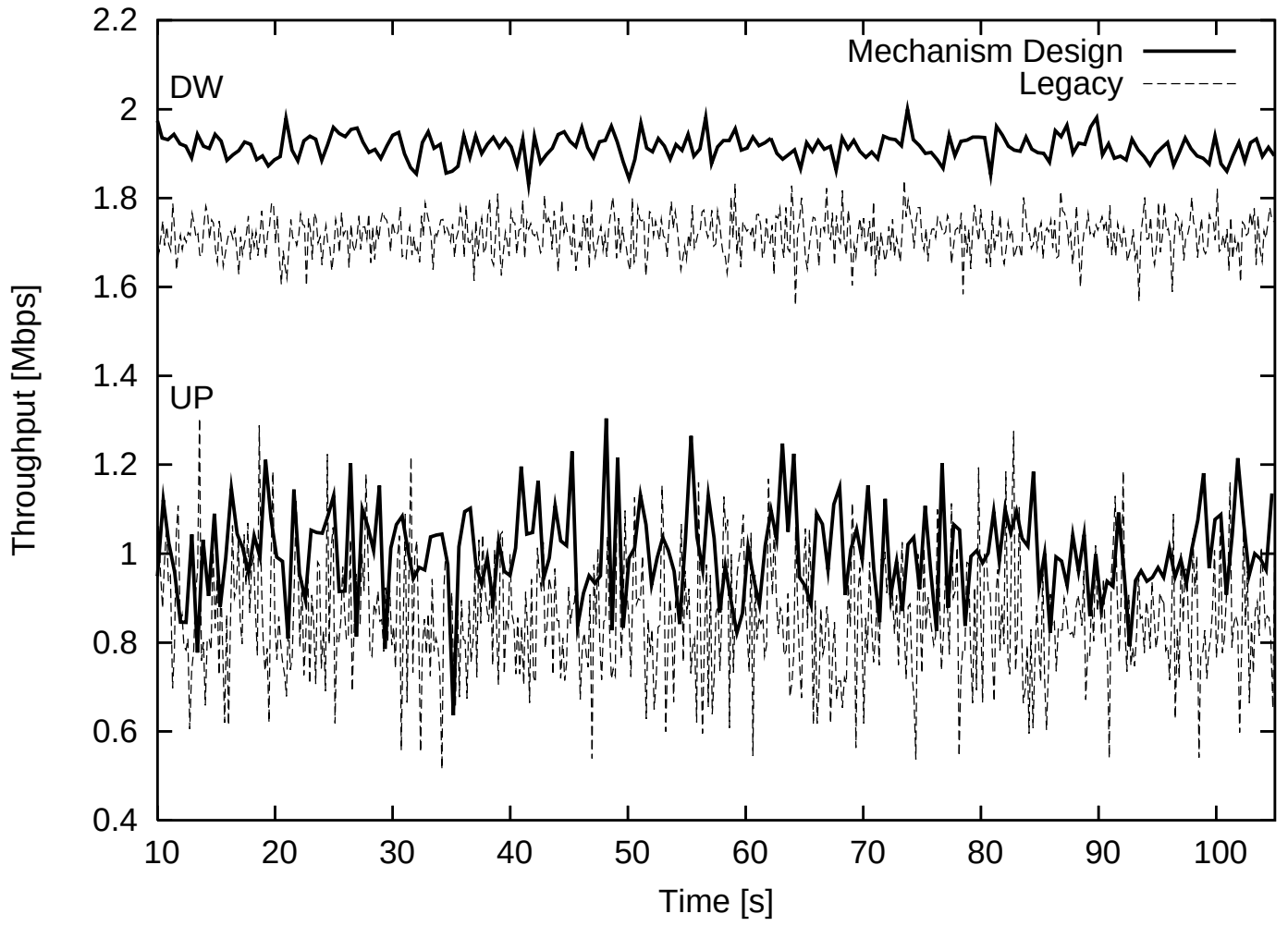

Fig. 10. Uplink and downlink throughput repartition at 11 Mbps under legacy AP (dashed lines) or mechanism design (bold lines), for n=10, k=0.5.

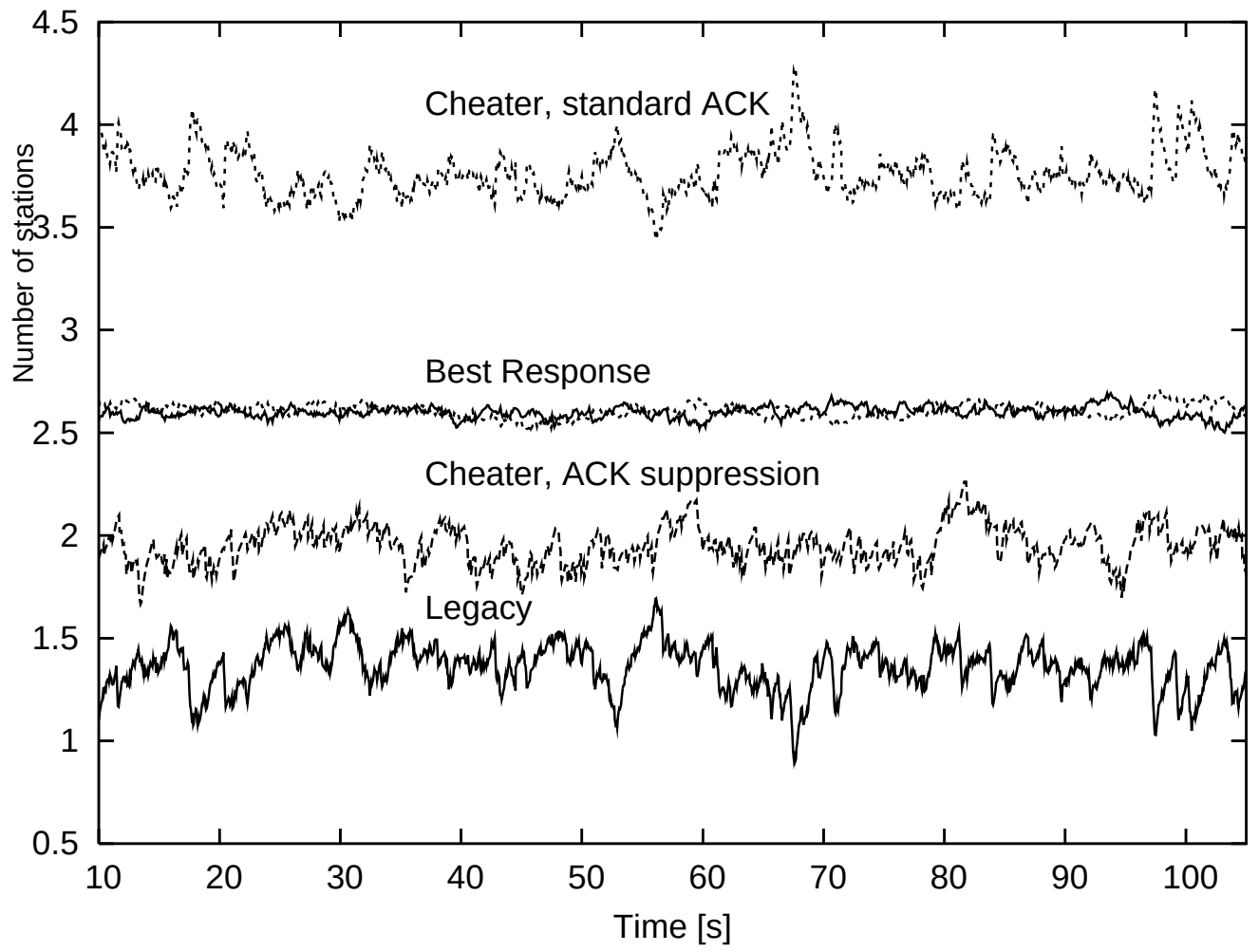

Fig. 11. Effects of the ACK suppression scheme on a cheater station throughput.

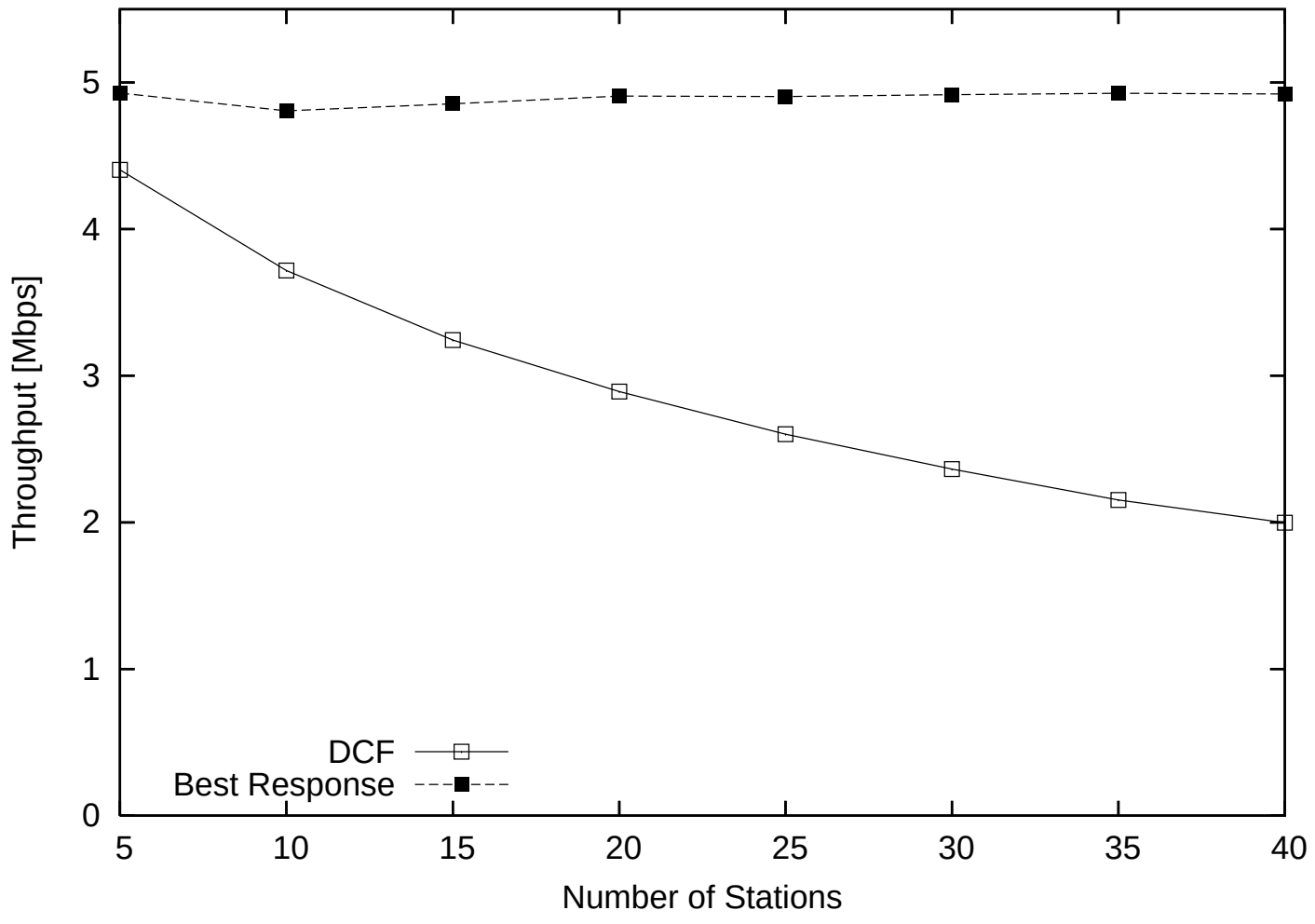

Fig. 12. Unidirectional case ($k = \infty$): aggregated throughput for various number of nodes. Best response is compared with standard DCF.